# REVIEWS

# The first decade of science with Chandra and XMM-Newton

Maria Santos-Lleo[1], Norbert Schartel[1], Harvey Tananbaum[2], Wallace Tucker[2] & Martin C. Weisskopf[3]

NASA's Chandra X-ray Observatory and the ESA's X-ray Multi-Mirror Mission (XMM-Newton) made their first observations ten years ago. The complementary capabilities of these observatories allow us to make high-resolution images and precisely measure the energy of cosmic X-rays. Less than 50 years after the first detection of an extrasolar X-ray source, these observatories have achieved an increase in sensitivity comparable to going from naked-eye observations to the most powerful optical telescopes over the past 400 years. We highlight some of the many discoveries made by Chandra and XMM-Newton that have transformed twenty-first century astronomy.

Typically, cosmic X-rays are produced in extreme conditions—from intense gravitational and magnetic fields around neutron stars and black holes, to intergalactic shocks in clusters of galaxies. Chandra[1] and XMM-Newton[2] have probed the space-time geometry around black holes, unveiled the importance of accreting supermassive black holes in the evolution of the most massive galaxies, demonstrated in a unique manner that dark matter exists, and confirmed the existence of dark energy. They have also tracked the production and dispersal of heavy elements by supernovae and measured the magnitude and rate of flaring of young Sun-like stars. Table 1 gives a subjective and by-no-means complete list of significant discoveries made using these observatories.

With an order-of-magnitude or more improvement in spectral and spatial resolution and sensitivity, Chandra and XMM-Newton have shed light on known problems, as well as opened new areas of research. These observatories have clarified the nature of X-ray radiation from comets[3], collected a wealth of data on the nature of X-ray emission from stars of all ages[4,5], and used spectra and images of supernova shock waves to confirm the basic gas dynamical model[6,7] of these objects. They have resolved into discrete point sources the diffuse emission from the plane of the Galaxy[8], as well as the diffuse extragalactic X-ray background[9]. They have discovered hundreds of supermassive black holes at the centres of galaxies and for many of those obtained high-resolution spectra that have

Table 1 | A sample of discoveries made using the Chandra and XMM-Newton observatories

| Topic | Discovery |
| --- | --- |
| Comets | Established charge-exchange as mechanism for X-ray emission. |
| Individual stars | Measured densities, temperatures and composition of hot plasmas, testing models for stellar evolution, X-ray emission from stellar coronae, and stellar winds. |
| Star formation and star-forming regions | Discovered X-ray emission from gas accreting onto stellar surfaces and influenced by magnetic fields; detected giant flares from young stars, with implications for planet formation. |
| Supernovae | Established that Kepler's supernova was a thermonuclear event. |
| Supernova remnants (SNRs) | Discovered a central compact object in the Cas A SNR and traced the distribution of elements indicating turbulent mixing along with an aspherical explosion. Imaged forward and reverse shock waves in several SNR, with implications for the acceleration of cosmic rays. |
| Pulsar wind nebulae | Resolved jets and rings of relativistic particles produced by young neutron stars. |
| Black hole accretion processes | Provided evidence for rotation of space-time around black holes; measured the efficiency of the accretion process; and detected jets and winds produced by black holes. |
| Galactic Centre | Measured the flaring of central black hole and resolved the galactic ridge emission into individual sources. |
| Starburst galaxies | Discovered evidence for enrichment of the interstellar medium and the intergalactic medium by starbursts. |
| Supermassive black holes and active galactic nuclei (AGNs) | Resolved the X-ray background radiation into discrete sources, mostly supermassive black holes; traced the history of supermassive black hole growth over cosmic timescales. |
| Active galactic nuclei feedback in galaxies and clusters of galaxies | Discovered evidence for heating of hot gas in galaxies and clusters by outbursts produced by supermassive black holes, supporting the concept that supermassive black holes can regulate the growth of galaxies. |
| Dark matter | Determined the amount of dark matter in galaxy clusters and, by extension, the Universe; observed the separation of dark matter from normal matter in the Bullet Cluster, demonstrating that alternative theories of gravity are very unlikely to explain the evidence for dark matter. |
| Dark energy | Observed galaxy clusters to generate two independent measurements of the accelerated expansion of the Universe. |

[1]XMM-Newton Science Operations Centre, European Space Agency, Villanueva de la Cañada, 28691 Madrid, Spain. [2]Chandra X-ray Center, Smithsonian Astrophysical Observatory, Cambridge, Massachusetts 02138, USA. [3]Space Science Office, NASA Marshall Space Flight Center, Huntsville, Alabama 35812, USA.

997

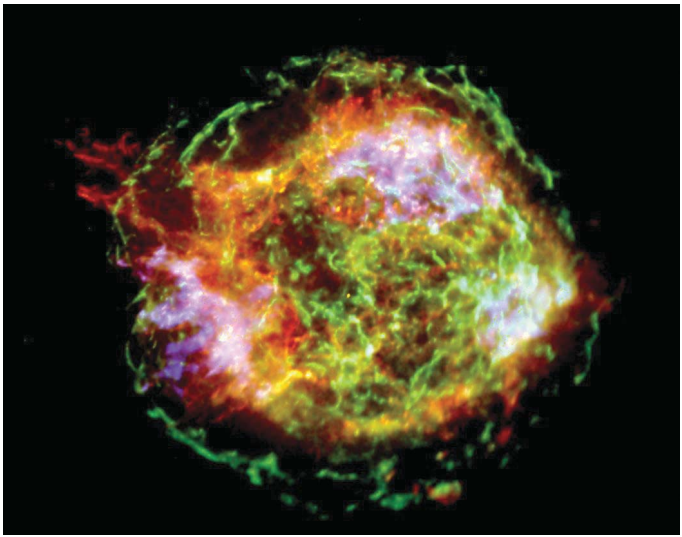

**Figure 1 | Chandra image of the Cassiopeia A supernova remnant.** Red, green and blue denote lower, medium and higher X-ray energies, respectively. Chandra's sub-arcsecond imaging clearly delineates the shock wave (bright green outer ring) generated by the supernova. A large jet-like structure (mostly red) protrudes beyond the shock wave towards the upper left. Reproduced with permission from ref. 89.

provided insight into the accretion process that powers the activity[10,11].

Within the Galaxy, surprises from Chandra and XMM-Newton included spectacular images of the Cassiopeia A (Cas A) supernova remnant (Fig. 1), of the Crab nebula (revealing rings and jets around the rapidly rotating neutron star[12]), and of numerous other pulsar wind nebulae[13]. The observatories found diverse X-ray emission mechanisms operating in normal stars[4]. Beyond the Galaxy, X-ray and radio data demonstrated convincingly that supermassive black holes and intense bursts of star formation can affect their environment (Fig. 2) on scales of hundreds of thousands of light years[14]. X-ray and optical data revealed the separation of dark and normal matter in a cosmic collision[15], and galaxy cluster surveys provided independent evidence for the accelerated expansion of the Universe[16,17].

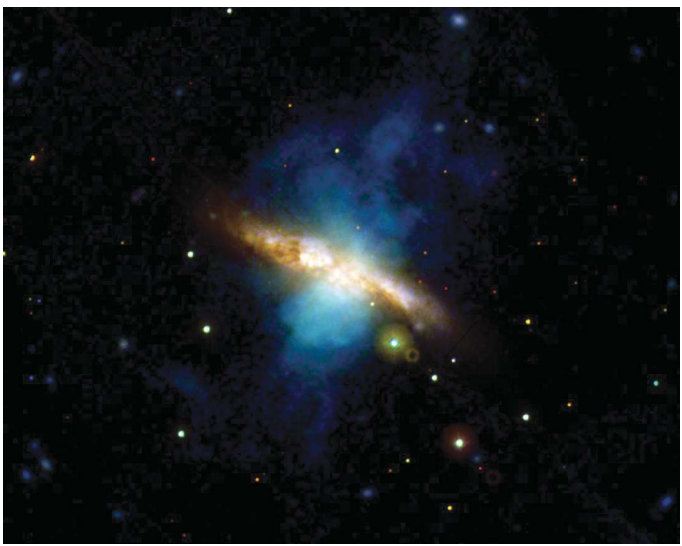

**Figure 2 | Composite of XMM-Newton optical-ultraviolet and X-ray images of the starburst galaxy Messier 82.** The optical-ultraviolet image—superimposed on the X-ray image (blue)—shows emission from the plane of this galaxy, obscuring dust lanes, and bright knots indicative of intense star formation. The X-ray image shows plumes of hot gas bursting from the galactic disk, in an extensive outflow driven by a burst of star formation. The X-ray emission is a combination of a continuum from many point sources in M82's central region and from collisionally ionized plasma whose temperature and abundance change along the outflow, plus emission lines due to charge-exchange reactions involving neutral metals (Mg and Si) in dust grains[90]. This composite image illustrates how simultaneous observations with XMM-Newton's on-board optical-ultraviolet camera enable extended wavelength coverage. Reproduced with permission from P. Rodriguez and the European Space Agency (ESA).

## Solar System objects

The interaction of energetic particles and radiation from the Sun with the atmospheres of planets, their satellites, and comets produces X-ray radiation through scattering, fluorescence, charge exchange, or the stimulation of auroral activity. Chandra and XMM-Newton have observed—and in many cases discovered—X-rays from Venus[18]; the Earth[19] and its moon[20]; Mars[21]; Jupiter[22,23], its aurorae[3] and disk[24], some of its moons[3,23] and the Io plasma torus[23]; Saturn[25–27] and its rings[28]; and numerous comets[29,30]. For detailed overviews of these observations, see references 3, 22 and 23.

An important mechanism for producing X-rays from Solar System objects is charge exchange, which occurs when a highly ionized atom in the solar wind collides with a neutral atom (gas or solid) and captures an electron, usually in an excited state. As the ion relaxes, it radiates an X-ray characteristic of the wind ion. Lines produced by charge exchange with solar wind ions such as C V, C VI, O VII, O VIII and Ne IX have all been detected with Chandra and XMM-Newton (for example, see Fig. 3).

The X-ray spectrum of a comet measures its out-gassing rate[29,30] and probes the solar wind *in situ*—in essence, a laboratory for highly ionized particles in low-density plasmas. Before XMM-Newton and Chandra, the origin of cometary X-rays was debated. Now, observational and theoretical work has demonstrated that charge-exchange collisions of highly charged solar wind ions with cometary neutral species are the best explanation for the emission[3].

Chandra and XMM-Newton have also discovered X-ray signatures of charge exchange in the exospheres of Venus and Mars, thus enabling measurement of planetary out-gassing rates, remotely and on a global scale. In addition, these observatories probe the planetary atmospheres through fluorescence and scattering of solar X-rays by constituents of the atmosphere (Fig. 3).

Planets with a magnetic field, such as Earth, Jupiter and Saturn, generate X-rays by auroral activity. Chandra discovered[23] time-variable X-ray flux from Jupiter that originates, not from the ultraviolet auroral zone as had been expected, but from higher latitudes near Jupiter's pole—regions that map to the outer boundary of Jupiter's magnetosphere. Charged particles precipitating into the Jovian atmosphere must therefore originate from much farther than the Io plasma torus. In addition, XMM-Newton observations[22] detected a component of the emission above 2 keV that is probably due to electron bremsstrahlung. Altogether these discoveries suggest a complicated current system in Jupiter's polar magnetosphere, requiring an important revision of the models.

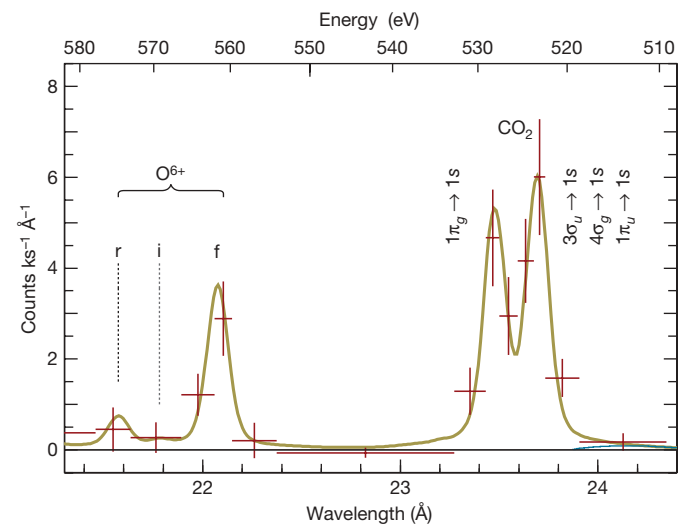

**Figure 3 | Portion of an XMM-Newton reflection-grating spectrum of the disk of Mars.** The spectrum shows prominent oxygen lines from two processes: charge exchange of solar-wind ionic oxygen (left) and fluorescence by solar X-rays of bound oxygen in the predominately $CO_2$ Martian atmosphere (right). Uncertainties are 1 standard deviation statistical errors and the solid line is a theoretical model. The symbols r, i and f refer to the resonance, intercombination and forbidden lines, respectively. Reproduced with permission from K. Dennerl (personal communication).

Chandra and XMM-Newton also detected charge-exchange X-rays from interaction of the solar wind with Earth's exosphere (geocorona), with material in and outside the magnetosheath[31], and possibly with the heliosphere[32]. The data suggest that at X-ray energies near 0.75 keV the total local soft X-ray intensity is from the heliosphere, whereas near 0.25 keV, the heliospheric component is significant, but does not explain all the observations. A mix of solar wind charge-exchange emission and a warm (~0.6 MK), rather than hot, local bubble in the interstellar medium seems to be needed[33]. This discovery affects the search for X-ray emission from the warm, hot, intergalactic medium (0.1 MK to 10 MK) plasma that has been proposed to contain over half the baryons at the current epoch. Ignoring the contribution of solar wind charge-exchange emission to the background could lead to the wrong interpretation of observed emission as due to the warm, hot, intergalactic medium[31].

### Individual stars and star-forming regions

Although the X-ray luminosity of stellar coronae is but a small fraction of the stellar bolometric luminosity, X-ray observations have proven to be good indicators of magnetic activity and the effectiveness of stellar magnetic dynamos. High-resolution X-ray spectroscopy with the objective gratings on XMM-Newton and Chandra have enabled measurement of many spectral lines—including resolved multiplets—and, in some cases, line profiles and line shifts. These measurements provide critical diagnostics of densities, temperatures and composition of hot coronal plasmas[4]. This capability has now enabled tests of models for X-ray emission from stellar coronae and, in the process, raised some intriguing questions. X-ray spectra of the coronae of cool stars have revealed an unexpected correlation between coronal abundance ratios and stellar magnetic activity and evolution. For example, the Ne/Fe ratio increases by an order of magnitude from the least active to most active coronae and even changes during a flare[4]. This effect is opposite to that seen in the solar corona, and is not yet completely understood.

Before XMM-Newton and Chandra, the mechanism for X-ray emission from massive, hot stars was generally thought to be due to shock waves colliding in the stellar wind, far above the stellar surface. The new data, although confirming some aspects of that model, show that adjustments are needed; for example, estimates for the mass-loss rate of massive stars may need to be revised downwards, or large-scale shocks from magnetically confined hot plasma may have a more important role than previously thought[4].

Two projects have been dedicated to study well-known, nearby, star-formation regions with many young stars: the XMM-Newton Extended Survey of the Taurus Molecular Cloud[34] (XEST) and the Chandra Orion Ultradeep Project[35,36] (COUP).

The Taurus molecular cloud is a test-bed for low-mass (~1 solar mass) star formation in dark clouds without luminous O-type stars. In this environment, stars form in isolation or in very small groups. The Taurus molecular cloud contains many T Tauri stars—very young stars that are still undergoing gravitational contraction and represent an intermediate stage between a protostar and a middle-aged star like our Sun.

Grating spectra from XEST observations have revealed that classical T Tauri stars accreting from circumstellar disks show excess flux in emission lines formed at low temperature—from, for example, C, N, O (Fig. 4)[4,37]. This 'soft X-ray excess' is thought to arise from magnetically collimated accretion columns at the stellar surface. Furthermore,

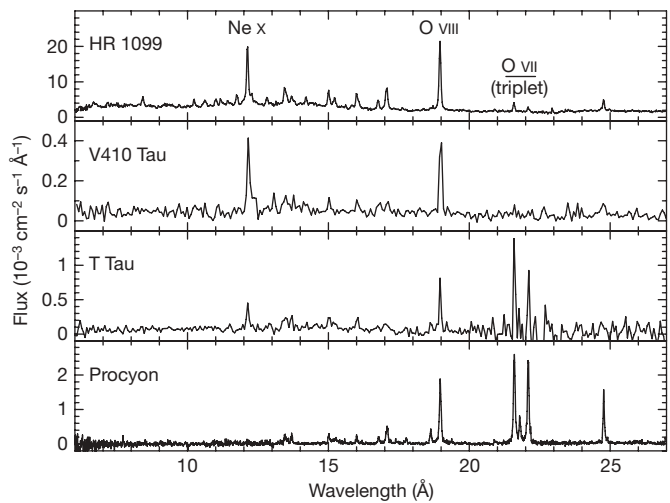

**Figure 4 | XMM-Newton X-ray spectra of different types of stars.** From top to bottom: the active binary HR 1099; the weak-line T Tauri V410 Tau; the classical T Tauri T Tau after absorption correction; and the inactive main sequence star Procyon. The O VII line, forming at 1–3 MK, is the strongest line in the X-ray spectrum of T Tau, despite the presence of large amounts of hot, coronal gas at 10–20 MK. Reproduced with permission from ref. 37.

the soft X-ray excess may be an important source of ionizing radiation for protoplanetary disks.

In contrast with the Taurus molecular cloud, the Orion nebula cluster is the prototype for 'clustered' star formation, where hundreds to thousands of low-mass stars form around massive O-type stars. The COUP image (Fig. 5) reveals X-ray emission from brown dwarfs, protostars with a wide range of ages, and stars ranging from solar-type to massive O-type. X-ray flares are seen at all stages, from protostars to older systems without circumstellar disks.

The COUP survey found that X-ray flares are ubiquitous in pre-main-sequence and young stars. Flares result from reconnection of multipolar surface magnetic fields (as on the Sun), but field lines reaching the protoplanetary disk may be involved[5,38]. Pre-main-sequence stars produce powerful flares that are about 100 times more intense and 100 times more frequent than solar flares[39]. X-rays from young-star flares may also dominate the ionization of a circumstellar protoplanetary disk, thereby altering its structure, dynamics and chemistry[40]. These processes may affect the formation of planets and their habitability.

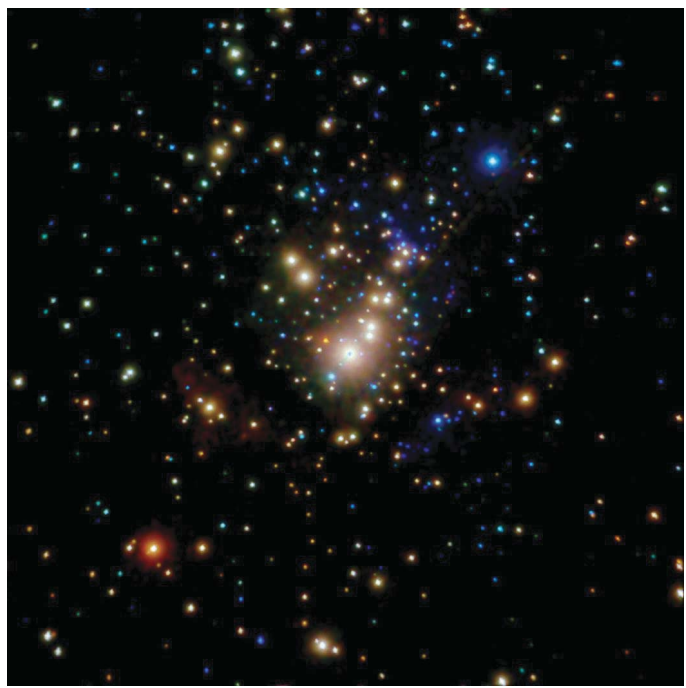

**Figure 5 | Deep Chandra image of the Orion nebula cluster.** This image shows more than a thousand young stars. Colours denote X-ray energy band: low (0.2–1.0 keV) in red, medium (1.0–2.0 keV) in green, and high (2.0–8.0 keV) in blue. Reproduced with permission from ref. 35.

The southwest side of Orion, a nearly circular region of much lower brightness than the northeast, is dominated by the massive stars. XMM-Newton observations revealed that this region is filled by hot ($\sim 2$ MK) plasma[41], produced by interaction of the hypersonic wind from the most luminous Orion nebula cluster star ($\theta^1$ Ori C) with the surrounding cold molecular cloud[42]. Detection of this plasma and similar structures in this and other regions with massive young stars reveals the large-scale effect of O-type star winds. Hot plasma from shocked winds leaking from the Orion nebula cluster feed a network of channels and bubbles of hot plasma in the Orion molecular cloud.

### Supernova remnants and massive stellar black holes

The final stage for a massive star is a supernova caused by collapse of the star's central core to form a neutron star (and then a black hole if the star is sufficiently massive). The enormous gravitational energy released during the collapse violently ejects most of the star at speeds of thousands of kilometres per second. The expanding supernova remnant (SNR) disperses heavy elements and generates shocks that heat ejected and ambient material, as well as accelerate particles to relativistic energies.

Chandra, Hubble Space Telescope and Spitzer observations of SN 1987A at different epochs trace the detailed evolution of this very young SNR (Fig. 6). X-rays are produced by shocks encountering dense fingers that protrude inward from a circumstellar ring of material. X-ray spectral line ratios measure the temperature and chemical abundance of shocked material[43], and line profiles give kinematic velocities of the gas[44]. Modelling of the Chandra images over a 10-year period shows clear evidence for the deceleration of the expansion as the remnant encounters circumstellar material[45].

Chandra and XMM-Newton have also enabled detailed studies of the temperature, chemical composition and structure of SNRs. For example, the Chandra image of the Cas-A SNR (Fig. 1) shows iron-rich ejecta outside silicon-rich ejecta, thus indicating that turbulent mixing and an aspherical explosion turned much of the original star 'inside out'[46]. Observations of Doppler-shifted emission lines for Cas A and other SNRs are providing three-dimensional information on the distribution and velocity of the supernova ejecta which will help to constrain models for the explosion[47]. The image of Cas A also revealed a point source that is probably a neutron star with a 10-cm-thick atmosphere of carbon[48].

In contrast to core-collapse supernovae of massive stars, a thermonuclear (type Ia) supernova occurs if a white-dwarf star becomes unstable by accreting too much matter or merging with another white dwarf. X-ray observations of the abundance and distribution of iron and oxygen in the Kepler SNR show that this remnant resulted from a type-Ia supernova. Kepler is unusual among such remnants in that it shows shock-heated circumstellar material, indicating that its progenitor lost substantial mass before exploding. Thus, Kepler is a rare example of a 'prompt' type-Ia supernova, which exploded about $10^8$ years after formation of its progenitor star[49]. This would require a progenitor star with a mass of $\sim 8$ solar masses that shed $\sim 7$ solar masses before becoming a white dwarf.

Chandra and XMM-Newton have also provided data which indicate that SNRs are important sites for the acceleration of cosmic rays[50,51]. The ability to distinguish the X-ray spectra of forward and reverse shocks has shown that X-ray emission from the forward shock is probably non-thermal synchrotron radiation from relativistic electrons accelerated in the forward shock. Non-thermal X-ray emission requires acceleration of electrons up to tens of teraelectronvolts and suggests acceleration of ions to even higher energies. (See ref. 51 for a discussion of other possible sources of cosmic rays in the Galaxy.) Observations of rapidly brightening and fading X-ray knots in SNRs RX J1713.7-3946 and Cas A imply short radiative lifetimes, strong magnetic fields and extremely efficient acceleration of electrons, and presumably ions, to high energies. Further evidence for the acceleration of ions in SNRs comes from the observed temperature behind the forward shock, which in some remnants is lower than expected from standard shock-wave theory. This implies that a significant portion of the post-shock energy is going into the acceleration of ions and electrons[50].

The shell-type SNRs are powered by the supernova itself. In contrast, filled-type (plerionic) SNRs—such as the Crab nebula (Fig. 7)—are powered by relativistic winds from young pulsars and ultimately their stored rotational energy. Chandra and XMM-Newton observations of these pulsar-wind nebulae have discovered[12] axisymmetric winds with shock fronts, jets and elongated nebulae[13].

Supernova remnants and pulsar wind nebulae will fade away after several thousand years, but if a neutron star or black hole is part of a binary star system, it may become an X-ray source once again as it accretes matter from its companion star. A long Chandra observation of the galaxy M33 provided a detailed study of the first eclipsing stellar-mass black-hole X-ray binary, M33-X7. This system contains one of the most massive (15.7 ± 1.4 solar masses) stellar black holes known and an unusually massive (70 solar masses) companion star. The system's properties—a massive black hole in a 3.45-day orbit, separated from its companion by only 42 times the radius of our Sun—are difficult to explain using conventional models for the evolution of massive stars, unless massive stars have significantly slower mass loss than usually thought[52]. Assuming that the black hole's spin is approximately aligned with the binary-system orbital angular momentum, analysis of spectral data combined from both observatories gave a precise measurement of the black hole's spin parameter[53]. In that an astrophysical black hole (being uncharged) can be described by just the mass and spin, this result yields a complete description of an asteroid-size object at a distance of about 3 million light years.

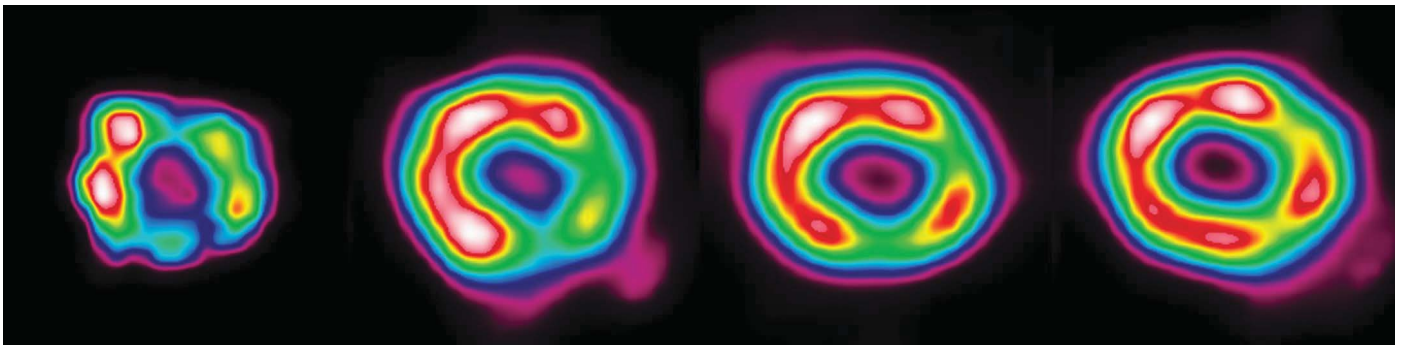

**Figure 6 | Chandra (2″ × 2″) images detailing the expanding and brightening ring surrounding the stellar explosion SN 1987A.** From left to right: January 2000, December 2002, August 2005 and October 2008. Reproduced with permission from K. Kowal Arcand and S. Park.

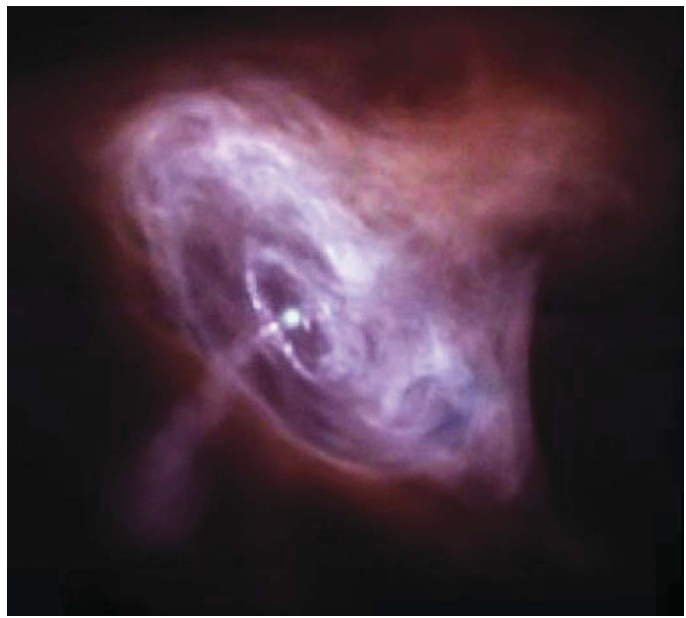

**Figure 7 | Chandra (2′ × 2′) image of the Crab nebula and pulsar.** The colours indicate intensity. Chandra provided the clearest view yet of the entire synchrotron nebula, discovering the inner ring and bright spot near the pulsar. Reproduced with permission from K. Kowal Arcand.

### Supermassive black holes

**Supermassive black hole census and evolution.** On a grander scale than the collapse of a single star, a supermassive black hole forms during the course of evolution of a galaxy through as yet poorly understood processes. A considerable body of evidence suggests that most (if not all) massive galaxies harbour a supermassive black hole. The supermassive black hole's mass scales with the mass of the host-galaxy bulge, indicating that co-evolution of the galaxy and its supermassive black hole occurs[14]. The supermassive black hole in the centre of our Galaxy, coinciding with the radio source Sagittarius A*, has a mass of only $\sim 3 \times 10^6$ solar masses and is extremely faint outside the radio band. However, Chandra and XMM-Newton have detected weak X-ray emission from Sagittarius A* marked with frequent, rapid X-ray flaring[54,55], which presumably results from clumps of material falling towards the supermassive black hole.

A supermassive black hole can grow through the gradual accretion of ambient material or by episodic swallowing of large clumps of matter—including stars and other supermassive black holes. For the galaxy RX J1242.6-1119A, combined X-ray (ROSAT (Röntgensatellit), XMM-Newton and Chandra) observations and optical and ultraviolet data showed that tidal disruption scenarios, whereby a nearby star is captured by the supermassive black hole, can explain the observed X-ray flaring over months or even years[56]. The Chandra image of NGC 6240, an ultraluminous infrared galaxy that resulted from the collision of two smaller galaxies, revealed that the galaxy's central region contains two active supermassive black holes that are destined to merge in a few 100 million years. This discovery provides direct evidence for a key stage of galaxy mergers and the growth of supermassive black holes[57].

Gravitational energy released through accretion into a supermassive black hole is generally thought to power active galactic nuclei (AGNs), which exhibit a wide range of luminosities from radio and Seyfert galaxies to quasars. Because most AGNs produce strong X-ray radiation that can penetrate the clouds of dust and gas shrouding many supermassive black holes, Chandra and XMM-Newton have proven to be valuable tools to find and study such systems.

Through deep exposures, these observatories are tracing the evolution of the supermassive black hole population, thus revealing the accretion history of the Universe[9,58]. Wide, moderately deep surveys have enabled astronomers to find elusive, highly obscured AGNs, and to study the population's evolution[59,60]. Figure 8 shows the XMM-Newton view of the Cosmic Evolution survey (COSMOS) field which is the deepest wide (>1 deg$^2$) extragalactic X-ray survey performed so far[61].

Among the most important findings is a strong cosmological evolution in the X-ray luminosity of AGNs[9,62–64]. Chandra and XMM-Newton data have yielded the first reliable space densities for X-ray-selected AGNs, covering a broad range in luminosity and redshift. The redshift measures the distance to an object and age of the Universe when the light was emitted: large redshifts imply large distances and young ages.

The evolutionary behaviour of AGNs depends upon X-ray luminosity: although the space density of high-luminosity AGNs peaks around a redshift $z \approx 2$, that of low-luminosity AGNs peaks at $z < 1$, confirming tendencies previously found with another X-ray satellite—ROSAT[62]. This drop occurs not because the accretion rate onto supermassive black holes evolves but rather because lower-mass supermassive black holes form or turn on more slowly than higher-mass supermassive black holes and because the latter switch off between $z = 1$ and the present. An explanation for this 'downsizing' is that high-luminosity AGNs are triggered in major mergers, whereas low-luminosity AGNs are due to more common non-merger events[65].

Combined X-ray spectral and timing observations also provide a probe of conditions near the event horizon of the supermassive black hole. XMM-Newton observations of the supermassive black hole in the galaxy MCG-6-30-15 have established that the broad iron line emission features observed can only be plausibly explained as lines shaped by relativistic effects[66]. In the galaxy NGC 3516, XMM-Newton observations of the modulation of a transient, redshifted iron K$\alpha$ emission line were used to track the orbit of a flaring hot spot just outside the event horizon for a supermassive black hole with a mass $\sim$30 million solar masses[67]. The combined spectral and timing data for iron L-line emission from the Seyfert galaxy 1H0707-495 also provides strong evidence that we are witnessing emission from matter within a fraction of a light minute from the event horizon of a rapidly spinning, massive black hole[68].

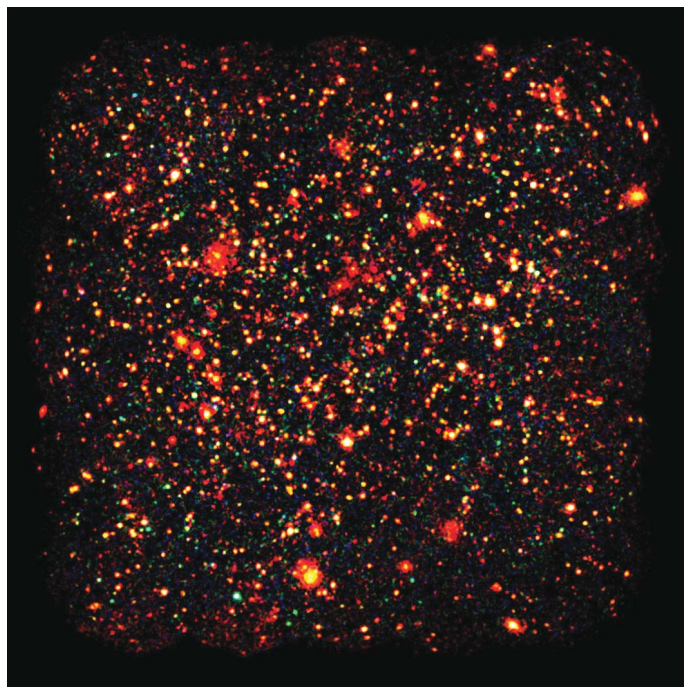

**Figure 8 | X-ray image of thousands of sources in the XMM-Newton COSMOS field.** The image covers three energy bands: 0.5–2 keV (red); 2–4.5 keV (green); and 4.5–10 keV (blue). Most of the $\sim$2,000 point-like sources are AGNs; more than 100 galaxy clusters appear extended. Reproduced with permission from G. Hasinger, N. Cappelluti and the XMM–COSMOS collaboration, and ESA.

**Supermassive black holes and clusters of galaxies.** Accretion onto a black hole is typically characterized by episodic flaring activity and possible quasi-periodic oscillations[69]. Accretion can also be accompanied by, or in some cases replaced by, conical winds of gas photo-ionized by radiation from the AGN, extreme outflows, and powerful, relativistic jets (for example, Fig. 9), which can represent a substantial fraction of the energy budget of an AGN[14,70]. Supermassive-black-hole-driven jets[71,72] seem to be the primary mechanism by which the supermassive black hole interacts with the gas in its host galaxy.

Direct evidence of AGN feedback comes from Chandra and XMM-Newton observations of clusters of galaxies, which have diameters as large as 10 million light years. In the absence of a heat source, the plasma would exhibit a 'cooling flow' with increasingly higher density and lower temperature in the central regions. Cooling-flow models that ignored heating experienced wide popularity with few dissenters for almost two decades. The cooling-flow debate took a dramatic turn when XMM-Newton and Chandra spectra of the cores of a number of 'cooling flow' clusters found that emission lines from iron L-shell ions—particularly Fe XVII—were much fainter than predicted by the models. This sets a strong upper limit on the amount of plasma with temperatures less than 10 MK (ref. 73) and indicates the presence of a heat source.

Evidence for a heat source was provided by a Chandra image[74] of the Perseus cluster of galaxies. Two vast X-ray-faint cavities discovered a decade earlier with ROSAT[75] were mapped in exquisite detail by Chandra. Radio data[75] show that these X-ray holes are filled with magnetized bubbles of (relativistic) synchrotron-emitting electrons, which displace the hot thermal plasma as they expand and rise buoyantly from the central black hole. Chandra images show quasi-spherical, ripple-like structures that, if interpreted as sound waves, can provide sufficient energy to offset the cooling.

A number of other galaxy clusters show this type of activity[76], most notably the relatively nearby Virgo cluster of galaxies, where the X-ray images reveal intricate details of supermassive-black-hole-generated outflow interacting with hot plasma[77,78]. The more distant cluster MS0735.6+7421 has huge X-ray cavities coincident with radio emission[76]. These cavities, which are 600,000 light years in diameter, indicate an outburst that has persisted for 100 million years and released energy of the order $10^{62}$ erg and indicate growth of the central black hole by as much as 100 million solar masses over that time.

The XMM-Newton and Chandra observations provide compelling evidence for a feedback scenario: gas cools in the cluster core and falls towards a central black hole. Accretion of the gas onto a spinning black hole then launches jets of relativistic particles that heat the cooling gas and stop the inflow, thereby shutting off the jets and allowing the cooling to resume. This picture also suggests an explanation for limits on the growth of massive galaxies at the centre of clusters[14].

Chandra and XMM-Newton have also provided new insights into the enrichment of the intracluster medium via supernova explosions. Many of the metals (elements from carbon to iron) in the intracluster medium are detected via their X-ray spectral signatures. Such data allow one to constrain models for supernova explosions[79] and to trace the iron content in clusters over cosmic time[80] with a decrease of nearly a factor of two from the present epoch to $z \approx 1$ when the Universe was about half its present age.

## Dark matter and dark energy

During the past decade there has been great progress in identifying and measuring the mass-energy components of the Universe. On the basis of a broad suite of astronomical observations, a concordance model has emerged in which only about 4% of the Universe is normal (baryonic) matter, 23% is dark matter, and 73% is dark energy[81]. X-ray observations have had an important role in compiling evidence for this conclusion.

In the merger cluster 1E 0657-558 (see Fig. 10) Chandra observations showed that the distribution of X-ray-emitting plasma is different from that of the total mass. This provides direct evidence for the existence of dark matter, thus rejecting a modification of the gravitational-force law[15]. The huge dark-matter accumulations in clusters of galaxies are laboratories for studying dark-matter particles. For example, the absence of decay lines in the XMM-Newton X-ray spectra of the

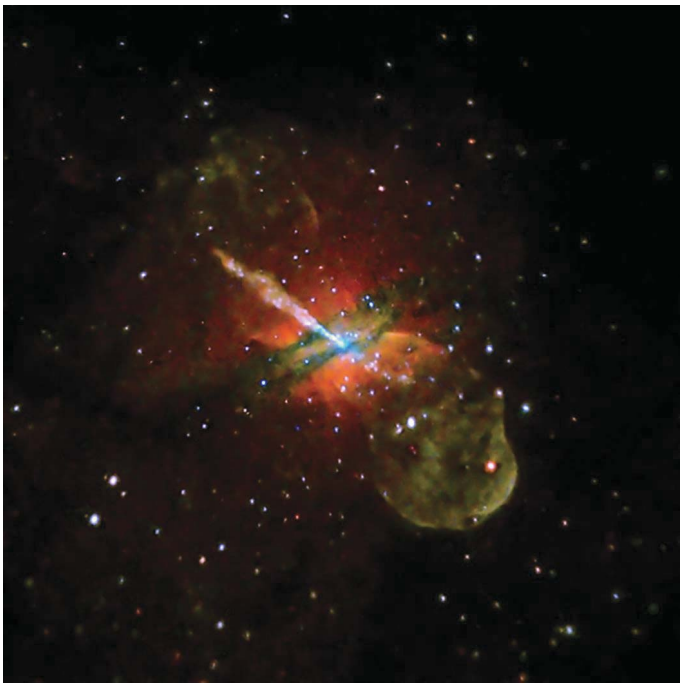

**Figure 9 | Chandra image of Centaurus A.** An ultra-deep X-ray view of the nearest galaxy with an active supermassive black hole. A prominent X-ray jet extends 13,000 light years to the upper left, with a shorter 'counterjet' in the opposite direction. Reproduced with permission of R. Kraft, NASA, the Chandra X-ray Center and the Center for Astrophysics.

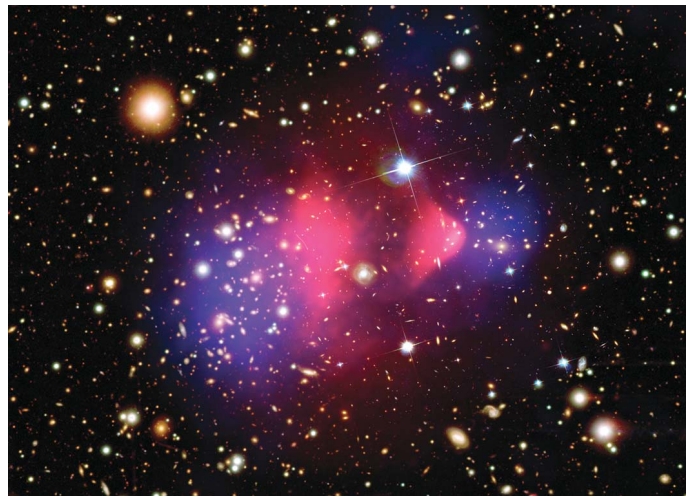

**Figure 10 | Colliding clusters of galaxies and dark matter.** Composite image of 1E0657-56 documents the collision of two large clusters of galaxies. Chandra resolved the X-ray emission (pink) produced by the vast clouds of hot gas which constitutes most of the baryonic matter. The optical image from Magellan and the Hubble Space Telescope shows the galaxies in orange and white. The blue areas indicate where most of the mass of the clusters lies, based upon gravitational weak lensing. Most of the matter (blue) is clearly separate from the normal matter (pink), directly showing that the mass in clusters is predominantly in some dark form that interacts only through gravity. Reproduced with permission from NASA, CXC and ref. 15.

Coma and Virgo clusters restricts the mass and mixing angle of hypothetical dark-matter sterile neutrinos[82].

Chandra and XMM-Newton data also showed that clusters of galaxies can be described by a universal mass profile[83,84]. This discovery constrains the dark-matter equation of state: self-interacting and adiabatic contracting dark matter can be eliminated[85,86].

On a larger scale, both the Hubble Space Telescope and XMM-Newton observed a 1.6-deg$^2$ region to map the three-dimensional structure of dark and baryonic matter. This large-scale structure comprises a loose network of filaments that grows over cosmic time. X-ray-emitting clusters of galaxies trace dark-matter concentrations at the intersections of these filaments[87].

Chandra observations of clusters of galaxies also provide a basis for precise determination of cosmological parameters in several new and unique ways. For example, the measured gas fraction $f_{gas}$ (where $f_{gas}$ = (mass in the X-ray-gas)/(total mass, including dark matter)) in clusters of galaxies depends on the assumed distance to a cluster, which in turn depends on the cosmological model. Moreover, for the largest, dynamically relaxed clusters, hydrodynamic simulations indicate that $f_{gas}$ should be nearly constant with redshift. Combining Chandra observations of $f_{gas}$ with this requirement yields constraints on the dark energy density and equation of state consistent and complementary with results obtained by other methods[16].

An independent technique for studying dark energy relies upon observing how the number of massive clusters per unit volume changes with time, compared with the predictions of cosmological models. With this 'growth of structure' method, Chandra data yield tight constraints on dark-matter and dark-energy content and on the dark-energy equation of state[17]. The data are consistent with the concordance cosmological model[81] wherein the dark energy arises from Einstein's cosmological constant.

## Future prospects

We cannot predict with certainty the major areas of discovery in the future and the impact that they will have. However, on the premise that "what's past is prologue"[88], we are convinced that deeper and wider surveys will yield a wealth of new discoveries. We expect that future investigations will address critical issues ranging from the subnuclear composition of a neutron star to the large-scale structure of the Universe. Targets of interest extend from the geocorona to deep surveys providing a complete census and evolutionary history of AGNs out to redshifts $z \approx 7$. Extended surveys of star clusters in the Galaxy and deep field observations of distant galaxies will provide insights into the processes that trigger star formation, and how bursts of star formation in the Galaxy as well as in the early Universe enrich the ambient gas with heavy elements and feed energy back into the gas. Similarly, long observations of individual galaxy clusters will probe and refine our understanding of the mechanism by which the energy from supermassive black hole feedback is coupled to the intracluster gas.

The scientific return from XMM-Newton and Chandra observations will continue to be enhanced by collaborations with ground-based and other space telescopes. For example, combined X-ray, infrared and millimetre spectral diagnostics using Herschel and the Atacama Large Millimeter Array will be crucial for probing accretion and outflow processes (and the role of magnetic fields) in proto-stars, young stars and proto-planetary disks. The search for the warm, hot, intergalactic medium, via absorption lines in X-ray grating spectra of relatively X-ray-bright blazars, will select targets based on data from optical surveys. The X-ray observations may confirm existing tentative detections or set limits on the amount of intergalactic highly ionized oxygen, which traces the 'missing baryons' in the local Universe. Dark-energy parameters will be better constrained by X-ray observations of galaxy clusters at higher redshifts, including those discovered in recently initiated microwave surveys—for example, the South Pole Telescope and the Atacama Cosmology Telescope, as well as the Planck mission. A key goal of the X-ray dark-energy research programme will be to test for a possible failure of general relativity through comparison of the effect of dark energy on cosmic expansion with its effect on the growth of cosmological structures.

**Acknowledgements** We feel privileged to be a part of the teams responsible for Chandra and XMM-Newton and wish to acknowledge both NASA and ESA for their continued support for these missions. We also acknowledge the contributions of the thousands throughout the world who have worked so hard to make these observatories so successful.

**Author Contributions** All of the authors worked together to produce the manuscript. M.C.W. led the effort on Solar System objects; M.S.-L. on individual stars and star-forming regions; M.C.W. on supernova remnants and massive stellar black holes; N.S. on supermassive black hole census and evolution; H.T. and W.T. on supermassive black holes and clusters of galaxies; and N.S. on dark matter and dark energy. W.T. served most ably as editor and supported all the sections. M.C.W. was the executive editor.

**Author Information** Reprints and permissions information is available at www.nature.com/reprints. The authors declare no competing financial interests. Correspondence should be addressed to M.C.W. (martin.c.weisskopf@nasa.gov).